\documentclass[aps, prmaterials, preprint, superscriptaddress]{revtex4-2}

\usepackage{graphicx}
\usepackage{dcolumn}
\usepackage{amsmath}
\usepackage{amssymb}
\usepackage{cleveref}
\usepackage{upgreek}

\crefname{figure}{Fig.}{Figs.}
\crefname{table}{Table}{Tables}

\begin{document}

\title{Thermal transport in ultrathin Si nanowires: a first principles study}



\affiliation{Department of Physics, University of Cagliari, Cittadella Universitaria, 09042 Monserrato (CA), Italy}
\affiliation{Sorbonne Unversit\'e, CNRS, Institut de Min\'eralogie, de Physiques des Mat\'eriaux et de Cosmochimie, IMPMC, UMR 7590, 4 place Jussieu, 75252 Paris Cedex 05, France}

\author{Konstanze R. Hahn}
\email{konstanze.hahn@dsf.unica.it}
\affiliation{Department of Physics, University of Cagliari, Cittadella Universitaria, 09042 Monserrato (CA), Italy}
\author{Claudio Melis}
\affiliation{Department of Physics, University of Cagliari, Cittadella Universitaria, 09042 Monserrato (CA), Italy}
\author{Fabio Bernardini}
\affiliation{Department of Physics, University of Cagliari, Cittadella Universitaria, 09042 Monserrato (CA), Italy}

\author{Lorenzo Paulatto}
\affiliation{Sorbonne Unversit\'e, CNRS, Institut de Min\'eralogie, de Physiques des Mat\'eriaux et de Cosmochimie, IMPMC, UMR 7590, 4 place Jussieu, 75252 Paris Cedex 05, France}

\author{Luciano Colombo}
\affiliation{Department of Physics, University of Cagliari, Cittadella Universitaria, 09042 Monserrato (CA), Italy}

\date{\today}

\begin{abstract}
Phonon properties of small Si nanowires in [110] direction have been analyzed using density functional perturbation theory. Several samples with varying diameters ranging from 0.38 to 1.5~nm have been calculated. It is found that the frequency of optical phonons at the zone center tend to decrease with increasing size of the nanowire.
Investigation of the phonon scattering rates has revealed very high values in the smallest sample which decrease with increasing nanowire size. A remarkable change in scattering rates is shown for increasing diameter from 0.53 and 0.78~nm to 0.86~nm. The higher phonon scattering could be attributed to an alignment of phonon modes at a specific frequency. 
Results of the thermal conductivity are lower with respect to bulk Si and are found between 15 and 102 W/mK. A trend of increasing thermal conductivity with increasing diameter can be observed. This effect is attributed to several changes in the phonon dispersion which are not necessarily correlated to the wire size. These explicit results have been compared to the thermal conductivity when boundary effect is approximated with Casimir scattering. The Casimir method substantially underestimates the results for explicit nanowires.

\end{abstract}

\maketitle 
\newpage
Silicon nanowires have been extensively investigated \cite{Rurali2010} since they represent an excellent test bed to study the fundamental physics of low-dimensional systems where confinement plays a major role. In addition, their abundance and physical properties make them  promising candidates for numerous next-generation technologies \cite{Peng2013,Mikolajick2013}. In this perspective, their thermal transport properties represent a key issue for many applications \cite{Zhang2013a}.
For instance, reducing the material dimensions to the nanoscale can lower the thermal conductivity and increase the thermoelectric efficiency. Control of thermal transport is of great importance as well in photovoltaic and microelectronic devices \cite{Browne2015, Wong2021, Cahill2003} where a high thermal conductivity is required. In addition, Si nanowires have also been studied as potential materials for thermal diodes and transistors \cite{Cartoixa2015}. 

In all applications a detailed description of the heat transport is crucial, in particular, for very small dimensions where bulk-like approximations are questionable. However, despite the large number of studies, no consensus has yet been reached on the thermal conductivity in ultrathin Si nanowires, neither of absolute values nor of the overall size-dependent behavior. 
Experimental studies have shown a decrease of thermal conductivity with decreasing nanowire size reporting  values from 10 to 40 W/mK for Si nanowires of size from 20 to 100~nm \cite{Li2003, Hochbaum2008, Poborchii2016}. A drastic reduction has been found down to 5 and 1.2 W/mK, respectively, if such nanowires are corrugated \cite{Poborchii2016} or have rough surfaces \cite{Hochbaum2008}.

From a theoretical point of view, several studies have been performed based on molecular dynamics (MD) simulations. Using the Green-Kubo formalism and the Stillinger-Weber potential, MD calculations have shown a steady decrease of the thermal conductivity down to about 1 W/mK with decreasing wire size from 5 to 1.6~nm at 300~K \cite{Volz1999}. Conflicting results, however, have been found: other authors using the same formalism but the Tersoff potential have reported a thermal conductivity larger than 45 W/mK and the opposite trend for a size reduction from 2 to 1.1~nm \cite{Donadio2010}.  
Similar size effects have been obtained using generalized tight-binding molecular dynamics simulations together with the Stillinger-Weber potential \cite{Ponomareva2007}. 
It is unclear, however, where the discrepancy between different studies using similar methods originates from. Volz et al. \cite{Volz1999} and Donadio et al. \cite{Donadio2010} used the Green-Kubo formalism but different interatomic force potentials. Instead, Ponomareva et al. \cite{Ponomareva2007} applied the same force field (Stillinger-Weber) as Volz et al. \cite{Volz1999}, but they calculated the thermal conductivity based on tight-binding simulations.

Different results have been obtained by Wang et al. \cite{Wang2010a} who used accurate phonon dispersions \cite{Mingo2003} to obtain the thermal conductivity. Their results show a steady decrease of thermal conductivity with decreasing size down to 1~nm and 0.33~W/mK similar to Volz et al. \cite{Volz1999} and good agreement with experimental results \cite{Kim2010} for larger nanowires.

\begin{figure}[htb]
\centering
    \includegraphics[width=0.8\textwidth]{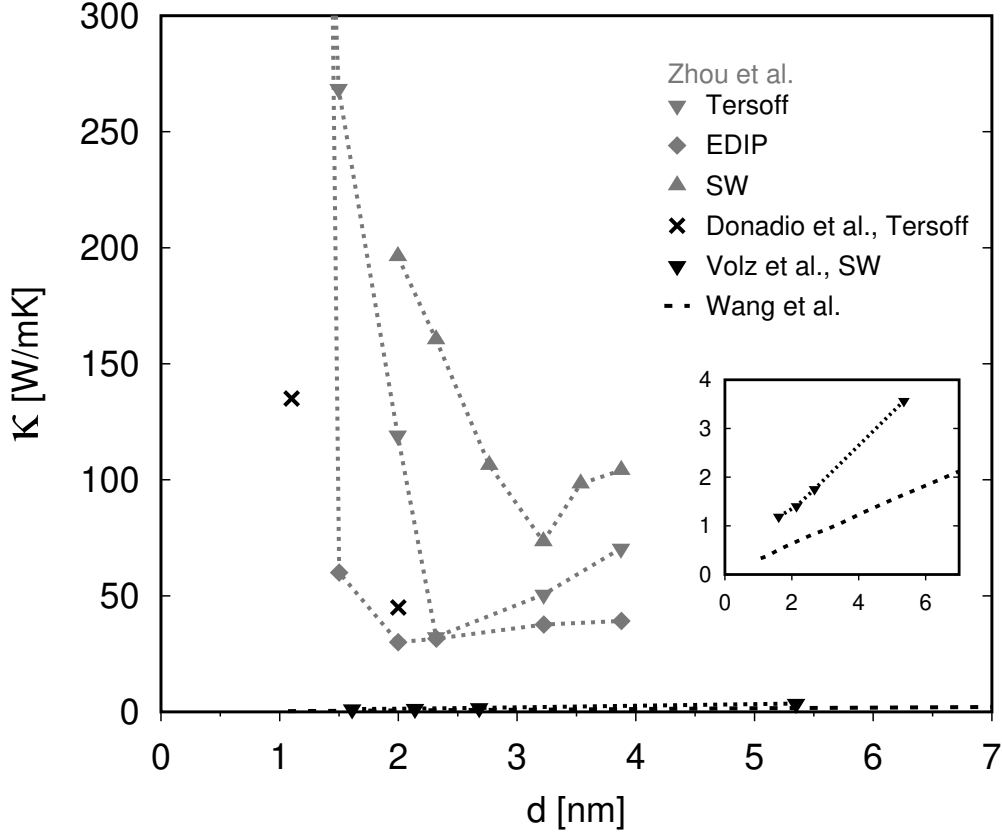}
        \caption{Thermal conductivity in small Si nanowires obtained by various computational approaches. Grey symbols are from MD simulations using the Green-Kubo approach and different force potentials \cite{Zhou2017}, black crosses have been obtained by Donadio et al. using a similar method and the Tersoff potential \cite{Donadio2010}. The black broken line (without symbols) shows data from Wang et al. where the thermal conductivity has been obtained from integration over all phonon branches \cite{Wang2010a} and black downward triangles resulted as well from MD calculations using the Stillinger-Weber potential \cite{Volz1999}. The inset is an amplification for better visualization of the results from references \cite{Wang2010a} and \cite{Volz1999}.
       }
        \label{fig:literature}
\end{figure}

This rather confusing scenario is graphically captured in \cref{fig:literature} and has been addressed recently by Zhou et al. \cite{Zhou2017}. Their study is again based on Green-Kubo MD simulations exploring different force fields. They find a similar trend of increasing thermal conductivity with decreasing size for very small nanowires in all of their calculations, but the absolute values vary remarkably (up to one order of magnitude) with the use of different force fields. With such huge differences the reliability of these methods is questionable, in particular, since the adopted potentials 
have previously shown to approximate other properties reasonably well \cite{Porter1997a}, however, leaning towards an overestimation of the thermal conductivity \cite{Schelling2002,Sellan2010,Fugallo2018}.

This state of affairs questions the reliability and accuracy of the results published so far. On the other hand, experimental data does not (yet) exist for Si nanowires of such small dimensions. Accordingly, a detailed study and understanding of the mechanisms directing the thermal conductivity in ultrathin nanowires is of great potential and can possibly predict structures with tailored thermal properties. 

Motivated by this, we have developed a detailed study on ultrathin Si nanowires, explored by first-principles calculations. Density functional theory (DFT) calculations have been applied to obtain the ground state of Si nanowires in [110] direction with diameters ranging from $d=$0.38 to 1.5~nm. Phonon dispersions have been calculated using density functional perturbation theory (DFPT). By further calculating the third-order interatomic force constants, anharmonic scattering terms have been evaluated, eventually leading to phonon scattering rates and the thermal conductivity for nanowires up to $d=$0.86~nm. 

Si nanowires with very small diameters have been generated using a lattice spacing of $a$=5.3803~\AA\ in the direction of the wire axis, corresponding to the value optimized for bulk Si with the computational settings used here \cite{Hahn2021,Hahn2021a}. The wire axis has been set to the [110] direction which has been shown to be favored in various fabrication processes, in particular for small sizes \cite{Li2003a,Wu2004,Zhang2005c}. 
Si atoms at the surface of the nanowire have been saturated with hydrogen atoms. 
The size of the generated samples range from 0.38 to 1.5~nm corresponding to a total of 14 (6 Si, 8 H) to 62 (42 Si, 20 H) atoms in the unit cell, respectively. Calculation of the thermal conductivity was computationally feasible only for samples with a diameter up to 0.86~nm (28 atoms). 
The cross-section of theses samples are shown as insets in \cref{fig:nw_scatt_varr}.

All calculations have been carried out with the \textsc{Quantum Espresso} program package \cite{Giannozzi2009, Giannozzi2017}, which, including its plug-ins, provides all elements necessary for the calculation of the third-order interatomic force constants (IFC) and related phonon scattering rates providing the lattice thermal conductivity  \cite{Fugallo2013,Paulatto2013,Ziman1960} 
\begin{equation}
\kappa_{\mathrm{L}}^{\alpha\beta}=\frac{\hbar^2}{N_0\Upomega k_B T^2}\sum\limits_{\mathbf{q}j}c_{\mathbf{q}j}^{\alpha}c_{\mathbf{q}j}^{\beta}\omega_{\mathbf{q}j}^2n_{\mathbf{q}j}\left(n_{\mathbf{q}j}+1\right)\tau_{\mathbf{q}j}
\end{equation}
where $c_{\mathbf{q}j}^{\alpha,\beta}$ are the branch-depending components of phonon group velocities in Cartesian coordinates $\alpha$ and $\beta$,  $\Upomega$ is the unit cell volume, $k_B$ is the Boltzmann constant and $\omega_{\mathbf{q}j}$,  $n_{\mathbf{q}j}$ and $\tau_{\mathbf{q}j}$ are the vibrational frequencies, the occupation and the relaxation time, respectively, of the corresponding phonon mode. 
The relaxation time $\tau_{\mathbf{q}j}$ is obtained as the inverse of the sum of all considered scattering terms. For the calculation of scattering terms in explicit nanowire samples, only anharmonic scattering rates are considered (boundary scattering is implied in the explicit geometry of the samples). A detailed description of the method has been provided previously \cite{Paulatto2013,Hahn2021a}. 
The volume $\Upomega$ has been adjusted in order to represent the volume of the nanowire without the vacuum present in the unit cell.

The local density approximation and the PZ (Perdew-Zunger) \cite{Perdew1981} functional have been applied to describe exchange and correlation interactions together with norm-conserving Trouiller-Martins type pseudopotentials \cite{Troullier1991} for the description of interactions between the frozen cores. 
Electronic ground state calculations have been carried out using a 1x1x8 \textbf{k}-point grid.
Phonon dispersion relations have been obtained using the \textsc{PH} module \cite{DalCorso1993} of \textsc{Quantum Espresso} with a 1x1x8 $q$-point grid in the direction of the axis of the nanowire which has been verified to be sufficient for converged results.
Calculation of the scattering mechanisms and the thermal conductivity has been carried out using the \textsc{D3Q} plug-in \cite{Paulatto2013} which provides the possibility to evaluate third-order interatomic force constants from density functional perturbation theory \cite{Baroni1987, Giannozzi1991, Baroni2001,Debernardi1994, Debernardi1995}.
For the determination of scattering terms, a 1x1x120 grid has been applied and the smearing parameter for energy conservation has been set to 1 cm$^{-1}$. 

In the one-dimensional structure of a nanowire there are four acoustic modes which can be assigned to one rotational mode, one longitudinal mode and two bending modes. The rotational mode, which corresponds to a torsion of the nanowire, and the longitudinal mode, which corresponds to a displacement along the wire axis, have a linear dispersion in the vicinity of $\Upgamma$. The other two modes, corresponding to displacements orthogonal to the wire axis (flexural modes), are quadratic close to $\Upgamma$. 

\begin{figure}[htb]
\centering
    \includegraphics[angle=270,width=0.75\textwidth]{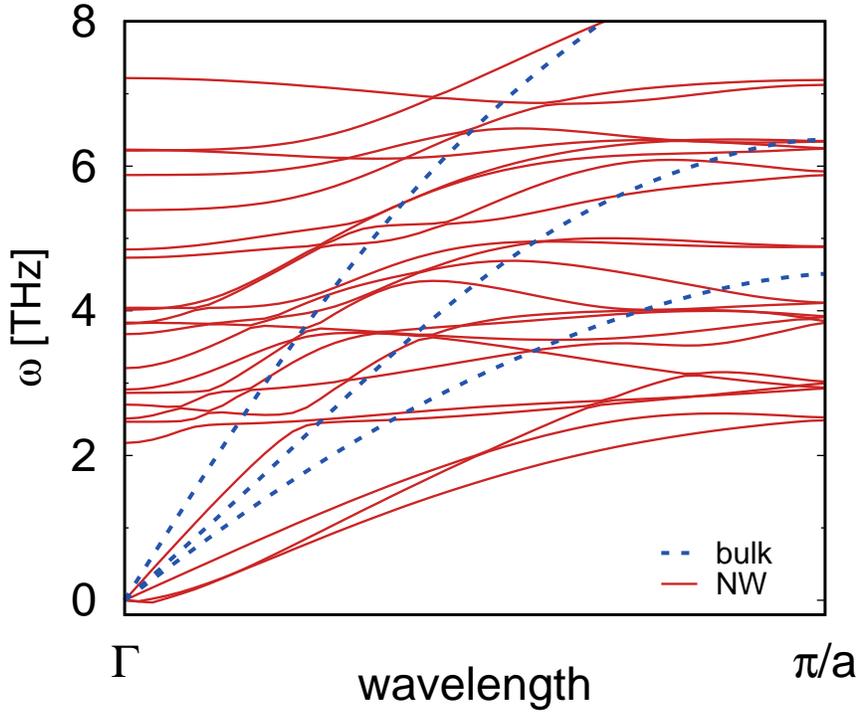}
        \caption{Phonon band structure of Si [110] nanowires (red lines) with $d=$0.86~nm in comparison to the phonon band structure in bulk Si (blue lines) in the corresponding direction.
       }
        \label{fig:phdos}
\end{figure}

\Cref{fig:phdos} shows the phonon band structure of a Si [110] nanowire with a diameter of 0.86~nm in comparison to the band structure in bulk Si in the corresponding direction. For better visualization we show phonon frequencies only up to 8 THz, keeping in mind that low frequency phonons are dominant in thermal transport in such systems.

Vibrational frequencies of acoustic modes in the nanowire are lower with respect to bulk Si and show good agreement to previous computational studies \cite{Peelaers2009,Zhou2017}.
Acoustic modes at the zone edge are two-fold degenerate at a frequency of 2.5 and 3 THz, respectively.

If dimensions are above a certain threshold (around 1-2~nm), it is commonly agreed that a reduction of characteristic dimensions of the material reduces its thermal conductivity. This effect has been observed for nanoparticles \cite{Melis2014b,Bux2009}, porous structures \cite{Lee2010}, thin films or superlattices \cite{Savic2013,Ferrando2015, Cheaito2012} and nanowires \cite{Volz1999,Li2003,Wang2010a,Liangruksa2011}. 
Recent theoretical studies, however, have shown the opposite trend for very small wire dimensions ($<2$~nm) \cite{Zhou2017, Gireesan2020} which has mainly been attributed to an increase in frequency of the lowest optical modes at the zone center with decreasing wire diameter which reduces the available channels for Umklapp scattering at low frequencies. 

\begin{figure}[htb]
\centering
    \includegraphics[width=0.8\textwidth]{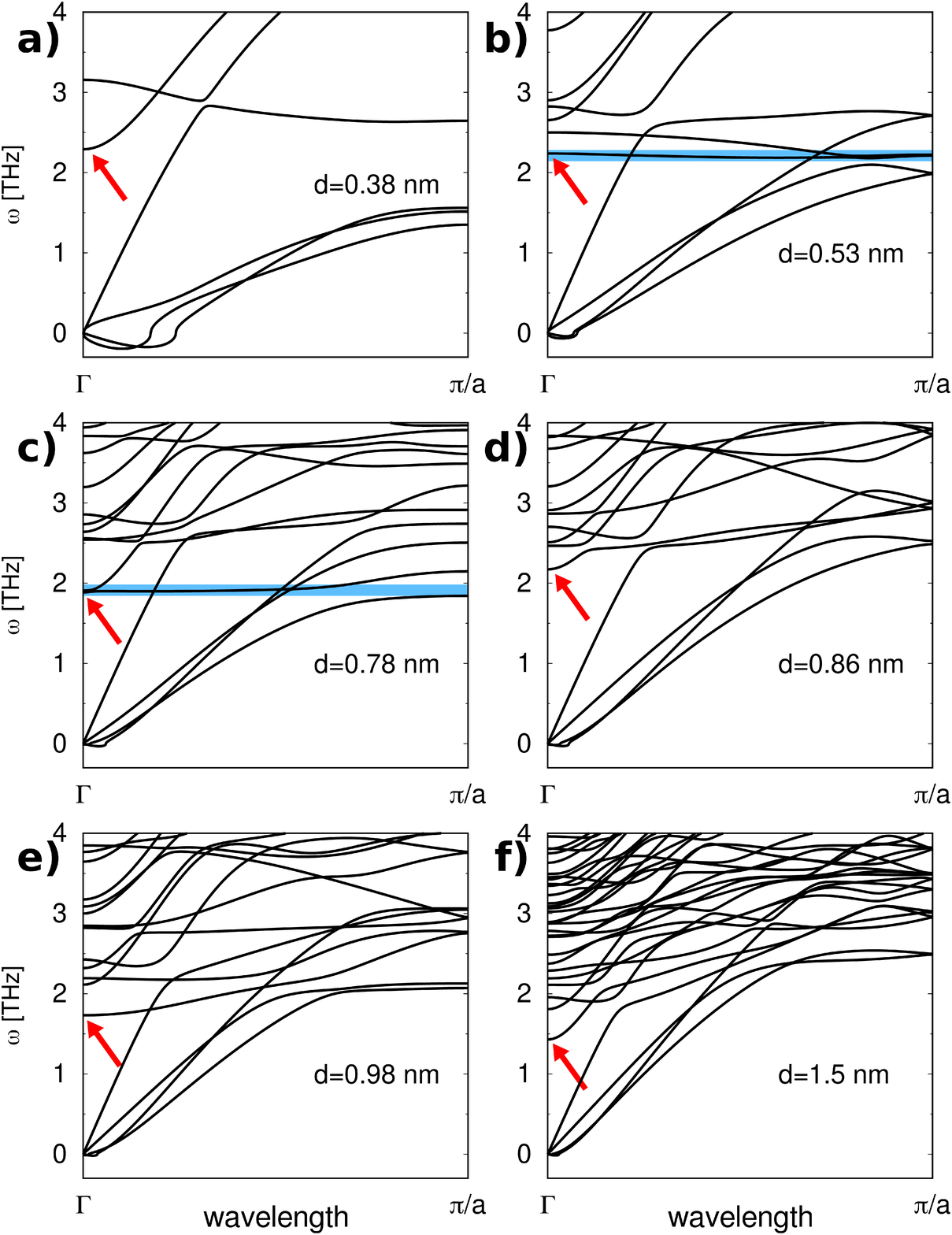}
        \caption{Phonon dispersion of Si nanowires in [110] direction with an average diameter ranging from 0.38 to 1.5~nm. Red arrows indicate optical phonons with the lowest energy at $\Upgamma$ and the blue horizontal line in b) and c) highlight the alignment of various phonon modes at the same frequency.
       }
        \label{fig:phdos_110_r}
\end{figure}

\begin{figure}[htb]
\centering
    \includegraphics[width=0.8\textwidth]{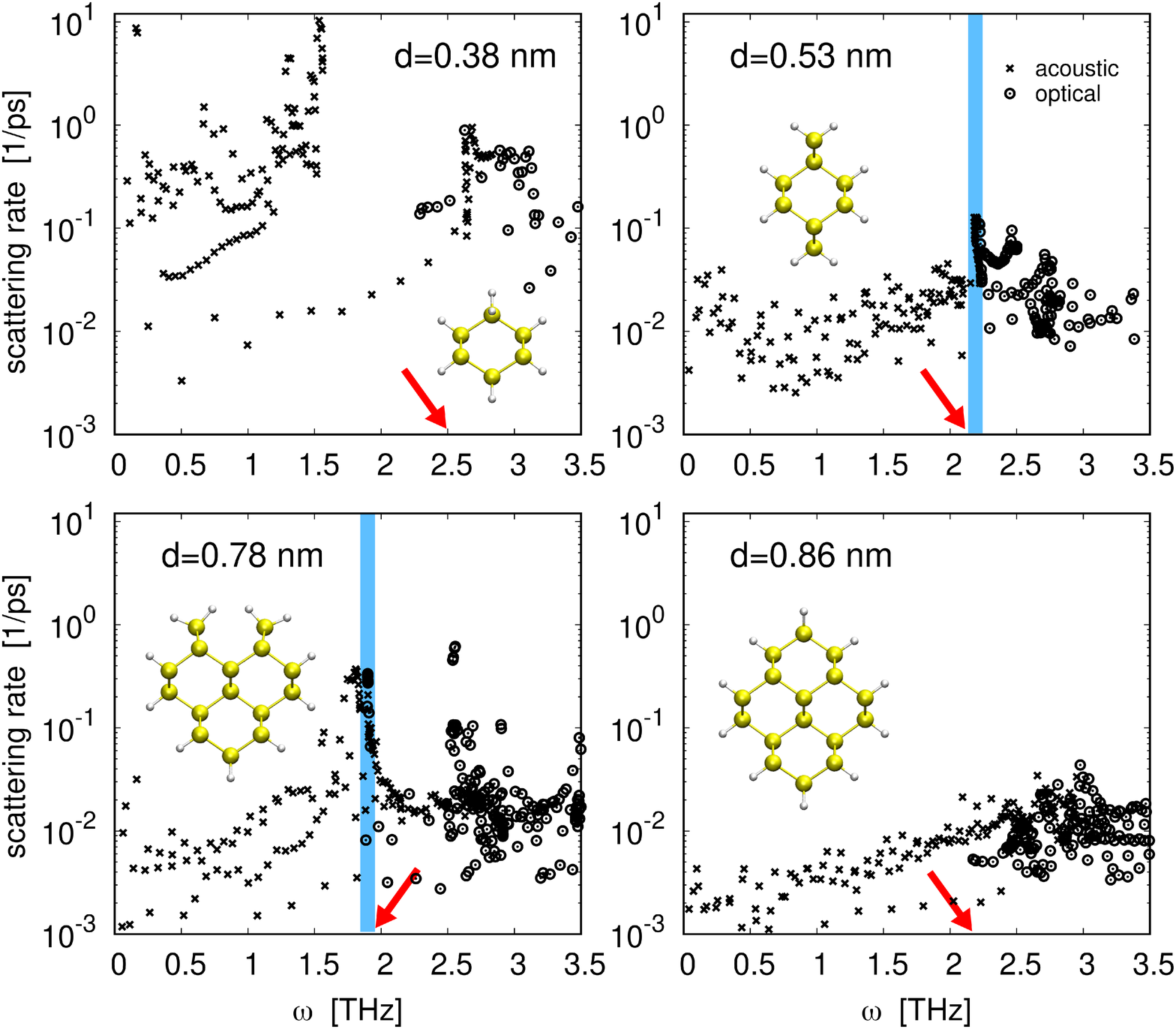}
        \caption{Phonon scattering rates of acoustic (crosses) and optical (circles) modes in Si [110] nanowires with an average diameter of  0.38~nm, 0.53~nm, 0.78~nm and 0.86~nm. Red arrows indicate optical phonons with the lowest energy at $\Upgamma$ and the vertical blue lines highlight the frequency where phonon modes shown notable alignment (see \cref{fig:phdos_110_r}). Insets show the cross-section of the corresponding sample.
       }
        \label{fig:nw_scatt_varr}
\end{figure}

This behavior in the phonon dispersion has been verified as shown in \cref{fig:phdos_110_r} for diameters from 0.38 to 1.5~nm. Red arrows indicate the position of the lowest optical frequency at the zone center. In particular, it is evident for an increase in diameter from 0.86 to 1.5~nm where the zone center optical frequency is found to decrease from 2.2 to 1.4 THz, respectively. It should be noted that negative frequencies in particular for the two smallest samples occur since the structures at the bulk lattice spacing of $a$=5.3803~\AA\ in wire direction is not fully relaxed. Ill-defined modes disappear when the lattice spacing in wire direction is increased. For a consistent comparison, however, we chose the bulk lattice spacing for all samples. Negative phonon modes have been ignored for the calculation of the thermal conductivity.

Explicit calculation of the phonon dispersion in nanowires and subsequent calculation of scattering terms using DFPT, as it has been done here, is computationally highly demanding. The number and diameter of the calculated samples had therefore been reduced to a feasible number with the largest diameter of 0.86~nm. 

It was expected to find less scattering channels with decreasing nanowire diameter as a result of the higher optical bands as has been described above based on recent studies \cite{Zhou2017, Gireesan2020}. 
However, the contrary has been found for decreasing diameters from 0.86 to 0.38~nm. An explanation can be derived from the phonon dispersion. As evidenced by the blue horizontal line for $d$=0.53 and 0.78~nm (\cref{fig:phdos_110_r}), the lowest optical band is almost dispersionless and crosses with acoustic modes at about 0.2~$\pi/a$. The alignment of different phonon modes at the same frequency and the crossing of acoustic and optical modes gives rise to enhanced scattering channels. This effect is demonstrated in \cref{fig:nw_scatt_varr} where the scattering rates are shown as a function of the phonon frequency for Si nanowires of different size. Enhanced scattering events are observed at this specific frequency (highlighted with a blue vertical line). In nanowires with $d$=0.53 and 0.78~nm this frequency is found at $\sim$2.2 and 1.8~THz, respectively.

Very high scattering rates (up to 11 1/ps) are present in the smallest sample with $d$=0.38~nm. No particular phonon frequency alignment of phonon modes or crossing of optical and acoustic modes is observed in contrast to the samples with $d$=0.53 and 0.78~nm. However, a peak in scattering is present around 1.5 THz which coincides with an overlap of acoustic modes at the zone border ($>0.5 \pi/a$). 

\begin{figure}[htb]
\centering
    \includegraphics[width=0.7\textwidth]{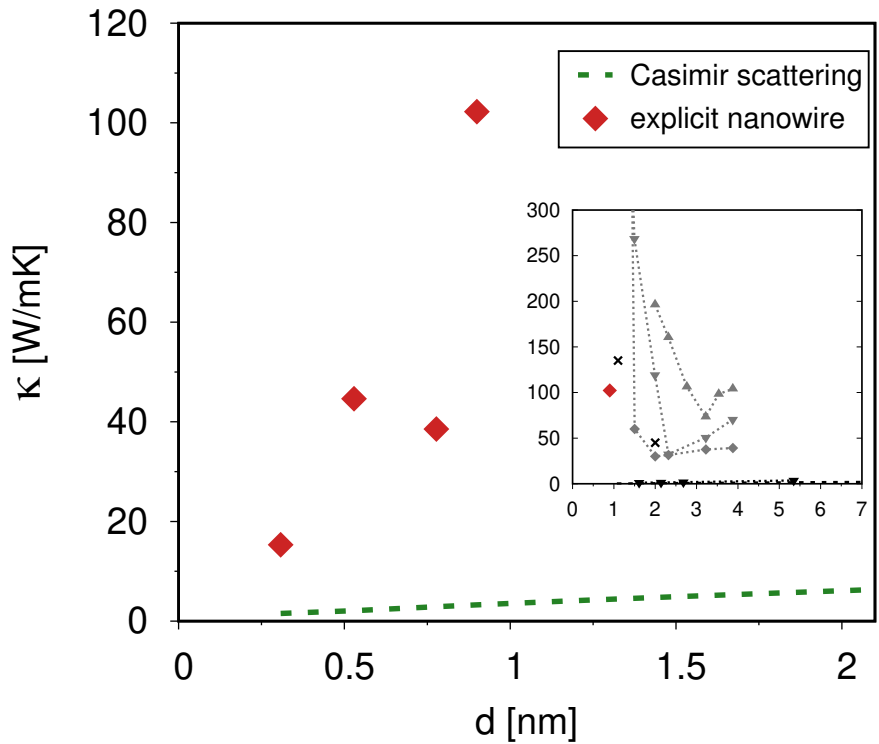}
        \caption{Thermal conductivity in ultrathin Si nanowires in [110] direction calculated explicitely (red diamonds) in comparison to results using the Casimir model (green broken line) where only boundary scattering is considered. The inset shows literature values in comparison to the largest sample studied here ($d$=0.86~nm). For legend and references of the inset see \cref{fig:literature}.
       }
        \label{fig:kappa}
\end{figure}

Based on the phonon dispersions and the scattering rates described above, the thermal conductivity has been calculated for diameters ranging from $d$= 0.38 to 0.86~nm. For these diameters, the thermal conductivity is found to increase from 15 to 102 W/mK (\cref{fig:kappa}).

This is in contrast to previous studies, that showed an increase in thermal conductivity with decreasing size in ultrathin nanowires \cite{Zhou2017, Gireesan2020}. These two studies attributed the behavior to a decrease in Umklapp scattering as a result of increasing optical bands with decreasing diameter as described above. The decrease in Umklapp scattering leads to dominating Normal scattering processes which promote hydrodynamic phonon flow, resulting in an increase in thermal conductivity. However, it should be noted that diameters of the nanowires calculated here are below the smallest size of the previous studies. In fact, the thermal conductivity of the 0.86~nm sample, is in reasonable agreement to Donadio et al. \cite{Donadio2010} (see inset of \cref{fig:kappa}). As has been discussed above, additional and increased scattering is found in the smallest samples of our study which supposedly dominate over Normal scattering events and thus suppressing hydrodynamic flow. Given our results of the thermal conductivity, we suggest that the largest sample ($d$=0.86~nm) lies in the regime of hydrodynamic flow while in all other samples additional scattering events are dominant thus reducing again the thermal conductivity. 
Larger systems, which could support this explanation, were not feasible for the calculation of the thermal conductivity within the DFPT method.

In order to estimate the effect of boundary scattering, the calculated thermal conductivity is compared to results using the Casimir model \cite{Casimir1938} (\cref{fig:kappa}). Within the Casimir approximation, anharmonic phonon scattering is obtained from the bulk phonon structure, while size effects are described as additional boundary scattering terms. 
As shown in \cref{fig:kappa}, the Casimir method notably underestimates the thermal conductivity calculated in explicit nanowires. 
This confirms that the sole consideration of boundary scattering is not capable of describing all effects that contribute to thermal transport in low-dimensional systems. Instead, an accurate representation of the phonon dispersion is necessary to be able to describe effects from anharmonic phonon interactions. 

In fact, the decrease of thermal conductivity for increasing diameter for 0.53 to 0.78~nm can be explained by the lower frequency of the dispersionless optical mode which gives rise to increased scattering channels (see \cref{fig:phdos_110_r}). 
On the other hand, the phonon dispersion in the $d$=0.86~nm sample does not show an alignment of modes at a certain frequency and the frequency of the lowest optical modes at the zone center is higher with respect to the 0.78~nm sample, indicating the presence of hydrodynamic flow and resulting in a remarkable increase in the thermal conductivity to 102 W/mK.

In summary, ultrathin Si nanowires and their phonon characteristics have been investigated using density functional perturbation theory. The phonon dispersion is discussed for nanowires in [110] direction with a size ranging from 0.38 to 1.5~nm. 
A decrease in frequency for optical modes at the zone center is observed with increasing size. 

In nanowires with $d=$0.53 and 0.78~nm, alignment of phonon modes at a certain frequency is observed which results in increased scattering channels at this frequency and subsequently to a decrease in thermal conductivity with respect to the sample with $d$=0.86~nm where such alignment is not present. 

As general trend it has been found that the thermal conductivity decreases with decreasing nanowire size, which can partly be attributed to such alignment of phonon modes. However, it is shown that small changes in the geometry of ultrathin nanowires can lead to drastic changes in the phonon dispersion which eventually alter the scattering mechanisms on various levels. For an accurate description of thermal transport properties in ultra-thin nanowires it is therefore crucial to rely on an explicit and correct description of the phonon dispersion. This conclusion implies that thermal properties in such ultrathin nanowires cannot be attributed solely to the wire size but are highly affected by the geometry and the resulting phonon dispersion.

\begin{acknowledgments}
This work is financed by Ministero dell'Universit\`a e Ricerca (MIUR) under the Piano Operativo Nationale 2014-2020 asse I, action I.2 "Mobilit\`a dei Ricercatori" (PON 2014-2020, AIM), through project AIM1809115-1.
\end{acknowledgments}

\end{document}